\documentclass[fleqn,twoside,10pt]{article}
\usepackage{espcrc2}
\usepackage{amsmath}
\usepackage{epsfig}
\usepackage{amssymb}
\usepackage{graphicx}

\input eqalign.sty     
\input macros.sty

\title{
\vskip -100pt{\large
\mbox{} \hfill DESY 03-151\\
\mbox{} \hfill SFB/CPP-03-42\\
\mbox{} \hfill September 2003\\}
\vskip 55pt
Non-perturbative renormalization of moments of parton
distribution functions
\thanks{Talk presented by A. Shindler. The work was supported by the
EU IHP Network on Hadron Phenomenolgy from Lattice QCD and by the
DFG under SFB/TR 09-03.}}
\author{A.~Shindler\address[NIC]{NIC/DESY Zeuthen,
Platanenallee 6, D-15738 Zeuthen, Germany},
M.~Guagnelli\address[ROMA]{Dipartimento di Fisica, Universit\`a di Roma 
        {\em Tor Vergata} and INFN, Sezione di Roma II,\\
	Via della Ricerca Scientifica 1, I-00133 Rome, Italy},
K.~Jansen\addressmark[NIC], 
F.~Palombi\address{E.~Fermi Research Center, c/o Compendio Viminale, pal.~F,
	I-00184 Rome, Italy},
R.~Petronzio\addressmark[ROMA], I.~Wetzorke\addressmark[NIC]
(ZeRo collaboration)}

\begin{document}

\begin{abstract}
We compute non-perturbatively the evolution of the twist-2 operators
corresponding to the average momentum of non-singlet quark densities. 
The calculation is based on a finite-size technique,
using the Schr\"odinger Functional, in quenched QCD. 
We find that a careful choice of the boundary conditions, 
is essential, for such operators, to render possible the computation.
As a by-product we apply the non-perturbatively computed renormalization constants
to available data of bare matrix elements between nucleon states.
\end{abstract}
\maketitle

\section{INTRODUCTION}
The accurate knowledge of hadron parton densities is an essential ingredient
for the experimental test of QCD at accelerator energies.
Their normalization is usually obtained from a fit to a set of reference
experiments and is used for predicting the behaviour of hard hadron processes
in different energy regimes.
The calculation of the normalization needs non-perturbative methods.
In order to have a phenomenological impact this determination must have a precision 
comparable with the experiments, and must have all the systematic 
uncertainties under control.
In this proceedings we will mainly summarize the results obtained in 
\cite{Gua03}, to which we refer for any unspecified notations.

\section{MOMENTS OF PARTON DISTRIBUTION FUNCTION}

The moments of parton distribution functions (PDF) 
are related to matrix elements of leading twist 
$\tau$ ($\tau=$dim-spin) operators of given spin, between hadron states $h(p)$
\bea
\hspace*{-5mm}
\langle h(p)|{\cal O}_{\mu_1 \ldots \mu_N}|h(p) \rangle &=& M^{(N-1)}(\mu)
p_{\mu_1} \cdots p_{\mu_n} \nn \\ 
&& + {\rm{terms~}} \delta_{\mu_i \mu_j} 
\eea
\be
\hspace*{-7mm}
\langle x^{(N-1)} \rangle (\mu) = M^{(N-1)}(\mu = Q) 
\ee
On the lattice the $O(4)$ symmetry is broken
to the hypercubic group $H(4)$, and the 2 irreducible 
representations of the non-singlet operators for $N=2$ are
\be
\hspace*{-7mm}
{\cal O}_{44} (x) = 
\bar\psi(x) \Big[ \gamma_4 \lrD_4 - \frac{1}{3} 
\sum_{k=1}^3 \gamma_k \lrD_k \Big] \frac{\tau^3}{2} \psi(x) \\
\label{eq:ope44}
\ee
\be
\hspace*{-9mm}
{\cal O}_{12} (x) = 
\bar\psi(x) \Big[ \gamma_1 \lrD_2 + \gamma_2 \lrD_1 \Big]
\frac{\tau^3}{2} \psi(x) 
\label{eq:ope12}
\ee


Our setup will be QCD in a finite space-time volume of size $T \times L^3$
with $T=L$. We choose the same boundary conditions of \cite{{theta},{Cap99}}, namely 
inhomogeneous Dirichlet boundary conditions at time $x_0 = 0$ and $x_0=T$ 
and periodic spatial boundary conditions up to a phase for the fermion fields.
\be
\psi(x + L \hat k)={\rm e}^{{\rm i}\theta_k}\psi(x)
\ee
\begin{figure}[t]
\begin{center}
\hskip-10mm
\epsfig{file=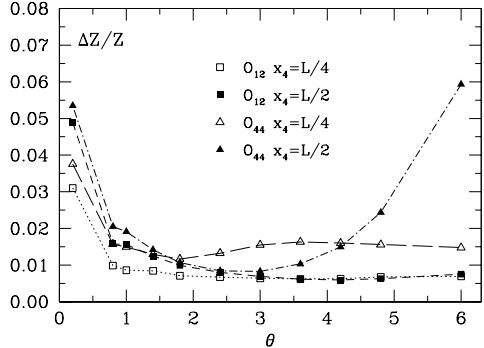,height=8cm,width=8cm,angle=360}
\vskip-40mm
\end{center}
\caption{Relative errors for the Z factor computed with 400 measurements on a $16^4$ lattice 
at $\bar g^2_{SF}(L)=3.48$}
\vspace{-0.4cm} 
\label{fig:rel_err}
\end{figure}
The strategy used to compute the non-perturbative evolution 
of the operators in eq. (\ref{eq:ope44},\ref{eq:ope12}) resembles the
strategy used by the ALPHA collaboration to compute the running quark mass \cite{Cap99}.
The evolution from initially large L (low $\mu$) to small L (high $\mu$)
is obtained applying the so called step scaling function (SSF) (cfr sect. 3).
Once the perturbative regime is reached (and this must be checked)
one continues the evolution in perturbation theory computing the 
(scale and scheme independent) RGI matrix element.
The connection with experiments is obtained then, 
making the adequate perturbative evolution of the RGI matrix element
in the $\MSbar$ scheme.

\section{RENORMALIZATION}

The renormalization conditions for the local operators are given by
\be
{\cal O}_R(\mu) = Z_{\cal O}^{-1}(a\mu) {\cal O}(a), \quad {\cal O}_R(\mu = L^{-1}) = {\cal O}^{(0)} \nn
\ee
The correlation functions to compute the Z factor are
\be
\hspace*{-9mm}
f_{\cal O}(x_0/L,\theta) = - {a^6 \over L^3}\;\sum_{\rm\bf x,y,z}\;
\langle {\cal O}(x)\;\Source_q({\rm\bf y},{\rm\bf z}) \rangle
\label{eq:corr_func}
\ee
\be
\hspace*{-9mm}
f_1(\theta) = - {a^{12} \over L^6}\;\sum_{\rm\bf u,v,y,z}\;
\langle \Source_q({\rm\bf y},{\rm\bf z}) \Source'_q({\rm\bf u},{\rm\bf v}) \rangle
\ee
where $\Source_q$ and $\Source_q'$ are suitable quark sources to probe the operators $\cO$.
With this definition the renormalization constants are obtained by
\be
Z(a/L,\mu) = c~ \frac{f_{\cal O}(x_0/L,\theta)}{\sqrt{f_1(\theta)}} ; \quad 
c = \frac{\sqrt{f_1^{(0)}(\theta)}}{f_{\cal O}^{(0)}(x_0/L,\theta)}. \nn
\ee


Optimal choice of $\theta$ and $x_0$ is mandatory to obtain a reliable signal of
the correlation function (\ref{eq:corr_func}).
In fig. \ref{fig:rel_err} a study of the relative error of the Z factor is performed.
A similar analysis can be performed for the cut-off effects and for
the convergence of perturbation theory computing the 2-loop anomalous dimensions for these
operators in the SF scheme \cite{Gua03}.
From these studies a good choice turns out to be
$\theta_1 = 1.0$, $\theta_{2,3} = 0$ and $x_0 = \frac{L}{2}$.

\begin{figure}[t]
\begin{center}
\hskip-10mm
\epsfig{file=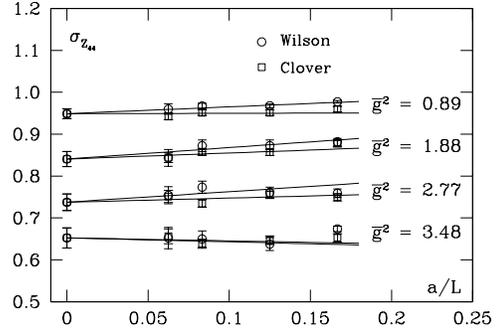,height=8cm,width=8cm}
\vskip-40mm
\end{center}
\caption{Continuum extrapolation of the SSF for selected values of $\bar g^2_{SF}(L)$}
\label{fig:cont_lim}
\end{figure}

To map out the $L$ dependence recursively we use the SSF, rigorously defined on the lattice by
\be
\hspace*{-9mm}
\sigma_{Z_{\cal O}} = \lim_{a \rightarrow 0} \Sigma_{Z_{\cal O}}(u,a/L)
\ee
\be
\hspace*{-9mm}
\Sigma_{Z_{\cal O}}(u,a/L) = \frac{Z_{\cal O}(u,2L/a)}{Z_{\cal O}(u,L/a)}, \quad u = \bar g^2_{SF}(L) 
\ee
The values of $\beta$ corresponding to a fixed running coupling are available in \cite{Cap99}. 
We have computed the SSF at $9$ values of the 
renormalized coupling ($\bar g^2_{SF}(L) = 0.8873$ to $3.48$) 
corresponding to a range of energies
that are roughly between $300$ MeV and $100$ GeV.
In order to have a better control on the continuum limit we have performed 
the computation with Wilson and Clover action, even if in both cases
one expects $O(a)$ lattice artefacts since the local operators are not improved.
In fig. \ref{fig:cont_lim} the continuum limit of the SSF
for some values of $\bar g^2_{SF}(L)$ is shown.
It is clear that a reliable (constrained) linear extrapolation with a small slope is possible. 

The formula that summarizes the whole strategy is given by
\bea
\langle \cO \rangle_{RGI} &=& \lim_{a\rightarrow 0}\frac{\langle \cO \rangle(a)}{Z_{\cal O}(a,\mu_0)} \times \nn \\
&& \sigma_{Z_{\cal O}}(\mu/\mu_0,\bar g^2(\mu)) {\cal F}_{SF}(\bar g^2(\mu)) \nn
\eea
where we use the $n=9$ SSF computed with $\mu=\mu_n$
\be
\sigma(\frac{\mu}{\mu_0},\bar g^2(\mu)) = 
\sigma(\frac{\mu_1}{\mu_0},\bar g^2(\mu_1)) \cdots \sigma(\frac{\mu_n}{\mu_{n-1}},\bar g^2(\mu_n)) \nn
\ee
to jump from the non-perturbative scale $\mu_0$ to the perturbative (ultraviolet)
scale $\mu$.
At this point one can try to do the perturbative matching using
\bea
{\cal F}_{SF}(\bar g^2(\mu)) &=& [\bar g^2(\mu)]^{-{\frac{\gamma_0}{2b_0}}} \times \nn \\ 
&& {\rm exp} \Big\{ -\int_0^{\bar g(\mu)} {\rm dx} \Big[ \frac{\gamma(x)}{\beta(x)} - \frac{\gamma_0}{2b_0}\Big]
\Big\} \nn
\eea
computed with 3-loop $\beta$ and 2-loop $\gamma$ functions.
If the perturbative matching has been successful the quantity 
\be
\hspace*{-9mm}
\mathfrak{\sigma}^{UV}_{INV}(\mu_0)=\sigma(\mu/\mu_0,\bar{g}^2(\mu))\;
{\cal F}_{SF}(\bar{g}^2(\mu))
\ee
should be independent from the ultraviolet scale $\mu$. Indeed what we find is that
the last 4-5 steps give a very nice plateaux (see fig. 10 in ref. \cite{Gua03}).

So it is possible to continue the evolution from the last step 
using perturbation theory. 
\begin{figure}[t]
\begin{center}
\hskip-10mm
\epsfig{file=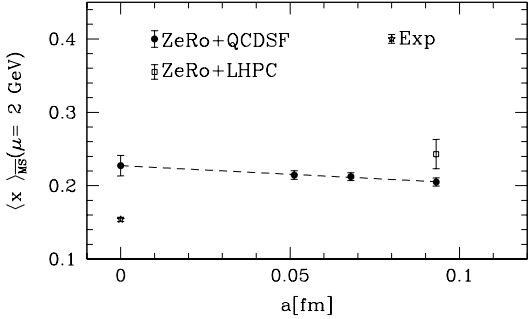,angle=360,height=8cm,width=8cm}
\vskip-40mm
\end{center}
\caption{Continuum limit of the non-perturbative renormalized first moment of the PDF
in a proton}
\vspace{-0.2cm} 
\label{fig:proton}
\end{figure}
\section{PRELIMINARY RESULTS}
Using perturbation theory is possible to compute the 
$Z^{RGI}$, that is the fundamental quantity to relate bare matrix elements
to any desirable scheme (e.g. $\MSbar$)

\be
\hspace*{-9mm}
\langle \cO \rangle_{RGI} = \lim_{a\rightarrow 0} \frac{\langle \cO \rangle(a)}{Z^{RGI}_{\cal O}(a)} = 
\langle \cO \rangle_{\MSbar}(\mu){\cal F}_{\MSbar}(\mu)
\ee
where
\be
Z^{RGI}_{\cal O}(a) = Z_{\cal O}(a,\mu_0) \frac{1}{\sigma(\mu/\mu_0,\bar g^2(\mu))}
\frac{1}{{\cal F}_{SF}(\bar g^2(\mu))} \nn
\ee
It is then clear that knowing $Z^{RGI}$ for a certain discretization
allows to compute in our case the parton average momentum $\langle x \rangle$ 
in the proton by
\be
\langle x \rangle_{\MSbar}(\mu) = \lim_{a\rightarrow 0} 
\frac{\langle x \rangle(a)}{Z^{RGI}_{\cal O}(a) {\cal F}_{\MSbar}(\mu)}
\ee


We then apply the $Z^{RGI}$ we have computed to the
unpublished data \cite{dataQCDSF} of QCDSF for the nucleon
bare matrix element and to the published data \cite{dataLHPC} of LHPC available
at only one value of $\beta$.
The continuum limit is shown in fig. \ref{fig:proton}.




\section{CONCLUSIONS}

We have computed in the continuum and in a fully non-perturbative way
the evolution of the twist 2 non-singlet operators with 
a very good precision (4\%).
This computation, combined with a calculation of the bare matrix element
between hadron states (for an application to pion matrix elements cfr. \cite{Ines03}),
gives the renormalization group invariant matrix element ${\langle x \rangle}^{RGI}$.
Then the RGI matrix element can be simply converted to any desirable scheme.
The precision of the experimental data requires a better control
on all the systematic uncertainties  (non-perturbative renormalization, continuum limit, 
chiral extrapolation, finite volume effects \cite{Ines03}, quenching).
In this contribution we have shown how to have complete control over the non-perturbative 
renormalization and on the continuum limit.
There is still a disagreement between the experiments and the lattice computation,
but there are also still systematic uncertainties in the lattice computation
that must be carefully analyzed.
It is clear that a comparison between experiment
and theory cannot be reliably done with a lattice simulation at one value
of the lattice spacing and without doing a non-perturbative
renormalization.
On the other systematic uncertainties works are in progress.

\def\NPB #1 #2 #3 {Nucl.~Phys.~{\bf#1} (#2)\ #3}
\def\NPBproc #1 #2 #3 {Nucl.~Phys.~B (Proc. Suppl.) {\bf#1} (#2)\ #3}
\def\PRD #1 #2 #3 {Phys.~Rev.~{\bf#1} (#2)\ #3}
\def\PLB #1 #2 #3 {Phys.~Lett.~{\bf#1} (#2)\ #3}
\def\PRL #1 #2 #3 {Phys.~Rev.~Lett.~{\bf#1} (#2)\ #3}
\def\PR  #1 #2 #3 {Phys.~Rep.~{\bf#1} (#2)\ #3}

\def\etal{{\it et al.}}
\def\ibid{{\it ibid}.}

\end{document}